\documentclass[12pt]{article}

\usepackage{amsmath,amsfonts,amssymb,latexsym}

\newtheorem{theorem}{Theorem}

\def\be{\begin{equation}}
\def\ee{\end{equation}}
\def\bq{\begin{eqnarray}}
\def\eq{\end{eqnarray}}
\def\beq{\begin{eqnarray*}}
\def\eeq{\end{eqnarray*}}
\def\f{\phi}
\def\c{\chi}
\def\a{\alpha}
\def\b{\beta}

\def\l{\lambda}
\def\m{\mu}

\def\na{\nabla}
\def\pa{\partial}

\begin{document}
\begin{titlepage}
\begin{flushright}
\end{flushright}

\vspace{0.7cm}

\begin{center}
{\huge Slice Energy and Conformal Frames in Theories of Gravitation}

\vspace{1cm}

{\large Spiros Cotsakis}\\

\vspace{0.7cm}

{\normalsize {\em Research Group of Cosmology, Geometry and Relativity}}\\
{\normalsize {\em Department of Information and Communication Systems Engineering}}\\
{\normalsize {\em University of the Aegean}}\\
{\normalsize {\em Karlovassi 83 200, Samos, Greece}}\\
{\normalsize {\em E-mail:} \texttt{skot@aegean.gr}}
\end{center}

\vspace{0.7cm}

\begin{abstract}
\noindent We examine and compare the behaviour of the scalar field slice
energy in different classes of theories of gravity, in particular
higher-order and scalar-tensor theories. We find a universal
formula for the energy and compare the resulting conservation laws
with those known in general relativity. This leads to a comparison
between the inflaton, the dilaton and other forms of scalar fields
present in these generalized theories. It also shows that all such
conformally-related, generalized theories of gravitation allow for
the energy on a slice to be invariably defined and its fundamental
properties be insensitive to conformal transformations.
\end{abstract}

\vspace{1cm}
\begin{center}
{\line(5,0){280}}
\end{center}

\end{titlepage}

\section{Introduction}
There have recently been many investigations on various structural
as well as evolutionary aspects of higher-order and scalar-tensor
theories of gravity (see Refs. \cite{rec1}-\cite{rec10} for
a partial list). While many of these analyses explore particular
problems in these modified frameworks, others aim to compare
such metric theories of gravitation from different points of view.
Although different these theories share two important common
characteristics: Firstly they can all be formulated in different
conformal frames.  It is well known (cf. \cite{ba-co88},
\cite{con5}) that the formulation of each of these modifications
and extensions of general relativity can be given in different,
conformally-related, spacetime manifolds, called conformal frames,
and depending on the particular problem one is working on, one
frame may prove more useful to all other conformally-related
ones. Secondly, they all require for their proper formulation in
at least one of the conformally-related frames the existence of
scalar fields. In some of these theories scalar fields are
mediators of the gravitational interaction while in others they emerge
as by-products of the transformation which relates two different
conformal frame representations of the same theory.

In fact these two characteristics turn out to be closely related:
the conformal transformation that relates two different conformal
representations of a theory is usually defined through the
introduction of a scalar field. Further the existence of
different conformal frames poses nontrivial relations between the
scalar fields present in them, which would otherwise have no
connection. It is therefore important to be able to state clearly
such relationships: What is the precise relation between the
scalar fields present in two different, conformally-related
frames? Are two frame representations of the same theory
mathematically and/or physically equivalent? Of course these
questions are not new and some of the related previous work  is
contained in Refs. \cite{brans,cot93,last}.

In this paper we analyze these questions from the viewpoint of a
geometric quantity, the energy of fields on a slice in spacetime,
and compare our findings about the behaviour of the slice energy
in such theories with that known in general relativity. This
comparison shows that slice energy is a kind of `universal
invariant' in metric theories of gravitation.  Further slice
energy may be used to clarify possible relations between the
different forms of scalar fields appearing in such theories, as
well as help uncover and compare the physical content in different
conformal frames in an invariant way.

In the next Section we write down the field equations which define
and describe the different theories we study and give the
conservation laws valid in each one of these frameworks to
establish our notation. Section 3 is the heart of this paper.
There we find how the slice energy behaves for the case of each
one of the theories given in Section 2. We conclude with a
discussion of how our results can be used to shed light on the two
aforementioned issues, namely the differences between the scalar
fields appearing in these theories and the possible physical
equivalence of these generalized modifications of general
relativity.

\section{Field equations}
We are interested below in a comparison of the conservation properties of
slice energy of certain fields in general relativity, higher-order gravity theories and
scalar-tensor theories of gravitation. We denote any matter field
present by the letter $\psi$. In general relativity we take the
field equations to be of the form
\be\label{gr}
G_{\a\b}=T_{\a\b}(\f)+T_{\a\b}(\psi ), \ee
where $G_{\a\b}$ is the
Einstein tensor, $\f$ is a scalar field with stress tensor
\be\label{t1}
T^{\a\b}(\f
)=\pa^\a\phi\pa^\b\phi-\frac{1}{2}g^{\a\b}(\pa^\l\phi\pa_\l\phi-2V(\phi)),
\ee
$T(\psi)$ represents the stress tensor of a field $\psi$, and
we assume the conservation identities $\na_\a T^{\a\b}(\f )= 0$
and $\na_\a T^{\a\b}(\psi)= 0.$

In higher-order gravity theories we consider the Jordan-frame
equations
\be\label{hog1} L_{\a\b}\equiv
f'R_{\a\b}-\frac{1}{2}g_{\a\b}f-\na_\a\na_\b
f'+g_{\a\b}\,\square\,_{g}f'=T_{\a\b}(\psi),
\ee which, because
$\na_a L^{\a\b}=0$, imply the conservation identities $\na_\a
T^{\a\b}(\psi)=0.$ The Einstein frame representation of this
theory is
\be\label{hog2}
\tilde{G}_{\a\b}=T_{\a\b}(\f)+\tilde{T}_{\a\b}(\tilde{\psi} ), \ee
where $\f =\ln f'$ and $T_{\a\b}(\f)$ is of the form (\ref{t1})
with $V(\phi)=(1/2)(f')^{-2}(Rf'-f)$, cf. \cite{ba-co88}. Here
the whole tensor in the right-hand-side is conserved,
\be\label{3.7} \tilde{\na}_\a\left(\tilde{ T}^{\a\b}(\f
)+\tilde{T}^{\a\b}(\tilde{\psi} )\right)=0, \ee but the two
components are not conserved separately, that is
\be
\tilde{\na}_\a \tilde{T}^{\a\b}(\f )\neq 0, \quad \tilde{\na}_\a
\tilde{T}^{\a\b}(\tilde{\psi} )\neq 0. \ee The field $\f$
appearing both in general relativity and in (the Einstein frame
representation of) higher-order gravity theories is in certain
contexts responsible for the existence of an inflationary period.
For concreteness we call it \emph{the inflaton} and
distinguish it from a scalar field, say  $\xi$, that may appear
directly in the Jordan frame equations (\ref{hog1}) in addition to
the matterfield $\psi$.

Lastly we take the defining equations of our scalar-tensor theory
to be the Brans-Dicke (BD) ones, with $\c$ denoting the BD scalar
field (everything we do below is valid if, instead of the BD
theory assumed here only for brevity, we consider the most general
scalar-tensor action having couplings of the form $h(\c)$, where
$h$ is any differentiable function of the field $\c$),
\be\label{st} S_{\a\b}\equiv\c\,
G_{\a\b}=T_{\a\b}(\c)+T_{\a\b}(\psi ). \ee The novel feature of
this equation is the requirement that, if in accordance with the
equivalence principle we assume that
\be
\na_\a T^{\a\b}(\psi)= 0,
\ee
\emph{only}, then, because $\na_\a G^{\a\b}=0$
we find
\be
\na_\a S^{\a\b}=\na_\a T^{\a\b}(\c).
\ee
Here $T^{\a\b}(\c)$ is not given by (\ref{t1}) but by a different, more
complicated, expression (cf. \cite{we72}, pp. 159-60). For definiteness
below we call the field $\c$ \emph{the dilaton} to distinguish it
from the other scalar fields appearing in the $f(R)$  Eqs. (\ref{hog1}), (\ref{hog2})
and in general relativity, Eq. (\ref{gr}). Many  currently popular
string theories appear as special cases of the scalar-tensor
equations.

\section{Slice energy}
Our starting point is the relation for the energies of a field on
two end-slices, $\mathcal{M}_{t_{1}}$ and $\mathcal{M}_{t_{0}}$, of
the time-oriented spacetime $(\mathcal{V},g)$ obtained in \cite{1}:
\be\label{1}
E_{t_{1}}-E_{t_{0}}=\int_{t_{0}}^{t_{1}}\int_{\mathcal{M}_{t}}
T^{\a\b}\na_{(\a}X_{\b)}d\mu
+\int_{t_{0}}^{t_{1}}\int_{\mathcal{M}_{t}}
X_{\b}\na_{\a}T^{\a\b}d\mu. \ee Here $\mathcal{V}=\mathcal{M}
\times \mathbb{R},$  $\mathcal{M}$ is a smooth manifold of
dimension $n$, $g$ a spacetime metric and the spatial slices
$\mathcal{M}_{t}\,(=\mathcal{M}\times \{t\})$ are spacelike
submanifolds endowed with the time-dependent spatial metric
$g_{t}$. (Greek indices run from $0$ to $n$, while Latin ones
from $1$ to $n$ and the metric signature is $(+-\cdots -)$.) For
$X$ any causal vectorfield of $\mathcal{V}$, we define the
\emph{energy-momentum vector $P$ of a stress tensor $T$ relative
to $X$} to be $P^{\b}=X_{\a}T^{\a\b}$ and the \emph{energy on
$\mathcal{M}_{t}$ with respect to $X$}, called hereafter the slice
energy, by the integral (when it exists)
$E_{t}=\int_{\mathcal{M}_{t}}P^{\a}n_{\a}d\m_{t},$ where $n$ is
the unit normal to $\mathcal{M}_{t}$ and $d\m_{t}$ is the volume
element with respect to the spatial metric $g_{t}$. We call
$P^{\a}n_{\a}$ the \emph{energy density} and assume that $X$ and
$T$ are smooth. Further, for the validity of the energy equation
(\ref{1}) on the spacetime slab
$\mathcal{D}=\Sigma\times[t_{0},t_{1}]$,
$\Sigma\subset\mathcal{M}$ and with $T$ having support on
$\mathcal{D}$, we take $\mathcal{M}$ to be compact or the field to
satisfy appropriate fall-off conditions at infinity. Fundamental
properties of the slice energy are proved in \cite{1}, Section 2.

In \cite{1} we showed how Eq. (\ref{1}) leads to relations
describing the energy exchange between the scalar field $\f$ and
the matter component $\psi$, in general but also especially in the
context of higher-order gravity.  Here we  follow a different
route: Starting from Eq. (\ref{1}) we derive relations showing the
dependence of the total slice energy of the system on the special
features of each one of the three theories given by Eqs.
(\ref{gr}), (\ref{hog1}) and (\ref{hog2}) and (\ref{st}). Writing
Eq. (\ref{1}) for the scalar field $\f$ and substituting from Eqs.
(\ref{gr}) and the conservation identity  for the terms $T(\f)$
and $X\na\, T$ respectively, we find \bq\label{gr-hog}
E_{t_{1}}(\f)-E_{t_{0}}(\f)&=&\int_{t_{0}}^{t_{1}}\int_{\mathcal{M}_{t}}
[G^{\a\b}-T^{\a\b}(\psi)]\,\na_{(\a}X_{\b)}d\mu \nonumber\\&=&
\int_{t_{0}}^{t_{1}}\int_{\mathcal{M}_{t}}G^{\a\b}\na_{\a}X_{\b}d\mu
-\int_{t_{0}}^{t_{1}}\int_{\mathcal{M}_{t}}T^{\a\b}(\psi)\na_{\a}X_{\b}d\mu.
\eq Using Stokes' theorem, the last term is just \be\label{12}
\int_{t_{0}}^{t_{1}}\int_{\mathcal{M}_{t}}T^{\a\b}(\psi)\na_{\a}X_{\b}d\mu=
\int_{\mathcal{M}_{t_{1}}}P^{\a}n_{\a}d\mu_{t_{1}}-
\int_{\mathcal{M}_{t_{0}}}P^{\a}n_{\a}d\mu_{t_{0}}
=E_{t_{1}}(\psi)-E_{t_{0}}(\psi), \ee and so, setting
$E_{t}(\f+\psi)=E_{t}(\f)+E_{t}(\psi)$, we find that in general
relativity the total slice energy of a system comprised of the
field $\f$ and a matter field $\psi$ depends on the Einstein
tensor as follows: \be\label{gr-e1}
E_{t_{1}}(\f+\psi)-E_{t_{0}}(\f+\psi)=
\int_{t_{0}}^{t_{1}}\int_{\mathcal{M}_{t}}G^{\a\b}\na_{\a}X_{\b}d\mu.
\ee Further, since $\na_\a G^{\a\b}=0$, integrating by parts and
using Stokes theorem we have
\be
\int_{t_{0}}^{t_{1}}\int_{\mathcal{M}_{t}}G^{\a\b}\na_{\a}X_{\b}d\mu=
\int_{\mathcal{M}_{t_{1}}}G^{\a\b}X_\a N_\b d\mu_{t_{1}}-
\int_{\mathcal{M}_{t_{0}}}G^{\a\b}X_\a N_\b d\mu_{t_{0}},
\ee
where $N$ is the unit normal to the slices. Using this form
we have the following result.

\begin{theorem}
The total slice energy of the system comprised of the scalar field
$\f$ and a matterfield $\psi$ satisfying the Einstein equations
(\ref{gr}), is given by
\be\label{gr-e2}
E_{t_{1}}(\f+\psi)-E_{t_{0}}(\f+\psi)=
\int_{\mathcal{M}_{t_{1}}}G^{\a\b}X_\a N_\b d\mu_{t_{1}}-
\int_{\mathcal{M}_{t_{0}}}G^{\a\b}X_\a N_\b d\mu_{t_{0}}.
\ee
In particular, when  $X$ is a Killing field of the metric $g$, the total slice
energy of the system  is conserved.
\end{theorem}
The terms of the form $\int_{\mathcal{M}_{t}}G^{\a\b}X_\a N_\b
d\mu_{t}$ represent a gravitational \emph{flux} through the slice
$\mathcal{M}_{t}$. When $X$ is a Killing field, the right hand side
of Eq. (\ref{gr-e1}) is zero and we have an integral conservation
law given by the equality of the two terms in the right hand side
of Eq. (\ref{gr-e2}) and this agrees with the corresponding result
originally given in \cite{synge}, Chap. VI.

The situation in higher-order gravity is in fact, despite the
different conservation laws, similar. In the Einstein frame we
have
\bq\label{gr-hog2}
E_{t_{1}}(\f)-E_{t_{0}}(\f)&=&\int_{t_{0}}^{t_{1}}\int_{\mathcal{M}_{t}}
[\tilde{G}^{\a\b}-\tilde{T}^{\a\b}(\tilde{\psi})]\,\tilde{\na}_{(\a}\tilde{X}_{\b)}d\tilde{\mu}
-\int_{t_{0}}^{t_{1}}\int_{\mathcal{M}_{t}}
\tilde{X}_{\b}\tilde{\na}_{\a}\tilde{T}^{\a\b}(\tilde{\psi})d\tilde{\mu}\nonumber\\&=&
\int_{t_{0}}^{t_{1}}\int_{\mathcal{M}_{t}}\tilde{G}^{\a\b}\tilde{\na}_{\a}\tilde{X}_{\b}
d\tilde{\mu}
-\int_{t_{0}}^{t_{1}}\int_{\mathcal{M}_{t}}\tilde{T}^{\a\b}(\tilde{\psi})
\tilde{\na}_{\a}\tilde{X}_{\b}d\tilde{\mu}\nonumber\\
&-&\int_{t_{0}}^{t_{1}}\int_{\mathcal{M}_{t}}\tilde{X}_{\b}\tilde{\na}_{\a}
\tilde{T}^{\a\b}(\tilde{\psi})d\tilde{\mu}.
\eq
Using Stokes' theorem, the middle term is
\bq
\int_{t_{0}}^{t_{1}}\int_{\mathcal{M}_{t}}\tilde{T}^{\a\b}(\tilde{\psi})
\tilde{\na}_{\a}\tilde{X}_{\b}d\tilde{\mu}&=&
\int_{\mathcal{M}_{t_{1}}}\tilde{P}^{\a}\tilde{n}_{\a}d\tilde{\mu}_{t_{1}}-
\int_{\mathcal{M}_{t_{0}}}\tilde{P}^{\a}\tilde{n}_{\a}d\tilde{\mu}_{t_{0}}\nonumber\\
&-&\int_{t_{0}}^{t_{1}}\int_{\mathcal{M}_{t}}\tilde{X}_{\b}\tilde{\na}_{\a}
\tilde{T}^{\a\b}(\tilde{\psi})d\tilde{\mu}\nonumber\\
&=&E_{t_{1}}(\tilde{\psi})-E_{t_{0}}(\tilde{\psi})
-\int_{t_{0}}^{t_{1}}\int_{\mathcal{M}_{t}}\tilde{X}_{\b}\tilde{\na}_{\a}
\tilde{T}^{\a\b}(\tilde{\psi})d\tilde{\mu}
\eq
and so, setting
$E_{t}(\f+\tilde{\psi})=E_{t}(\f)+E_{t}(\tilde{\psi})$, we find that in
higher-order gravity, because of the marvelous fact that the terms of the
general form $\int \tilde{X}\tilde{\na} \tilde{T}(\tilde{\psi})$
which were absent in general relativity now precisely cancel each
other, the total slice energy of a system composed of the field
$\f$ and a matter field $\tilde{\psi}$ in the Einstein frame
depends on the Einstein tensor in the same way as before:
\be\label{hog-e1}
E_{t_{1}}(\f+\tilde{\psi})-E_{t_{0}}(\f+\tilde{\psi})=
\int_{t_{0}}^{t_{1}}\int_{\mathcal{M}_{t}}\tilde{G}^{\a\b}\tilde{\na}_{\a}\tilde{X}_{\b}
d\tilde{\mu}.
\ee
Hence we arrive at  the following result.

\begin{theorem}
The total slice energy of the system composed of the scalar field
$\f$ and a matterfield $\tilde{\psi}$ satisfying the Einstein equations
(\ref{hog2}), is given by
\be\label{hog-e2}
E_{t_{1}}(\f+\tilde{\psi})-E_{t_{0}}(\f+\tilde{\psi})=
\int_{\mathcal{M}_{t_{1}}}\tilde{G}^{\a\b}\tilde{X}_\a \tilde{N}_\b d\tilde{\mu}_{t_{1}}-
\int_{\mathcal{M}_{t_{0}}}\tilde{G}^{\a\b}\tilde{X}_\a \tilde{N}_\b d\tilde{\mu}_{t_{0}}.
\ee
In particular, when  $X$ is a Killing field of the metric $\tilde{g}$, the total slice
energy of the system  is conserved.
\end{theorem}
Note that if we have a scalar field $\xi$ in addition to the matter field
$\psi$ present in the original Jordan frame of the higher order
gravity theory, then we obtain a  result similar to that in
general relativity but with $L_{\a\b}$ in place of the Einstein
tensor, namely,
\be\label{hog-e3}
E_{t_{1}}(\xi+\psi)-E_{t_{0}}(\xi+\psi)=
\int_{\mathcal{M}_{t_{1}}}L^{\a\b} X_\a N_\b d\mu_{t_{1}}-
\int_{\mathcal{M}_{t_{0}}}L^{\a\b} X_\a N_\b d\mu_{t_{0}}.
\ee
Then  terms of the form $\int_{\mathcal{M}_{t}}L^{\a\b}X_\a N_\b
d\mu_{t}$  represent a \emph{higher-order gravitational flux} through the slice
$\mathcal{M}_{t}$. When $X$ is a Killing field,  we again have an integral
conservation law as before.

We now move to the analysis of the scalar-tensor theory
(\ref{st}).  In this case Eq. (\ref{1}) becomes
\be\label{st-e1}
E_{t_{1}}(\c )-E_{t_{0}}(\c )=\int_{t_{0}}^{t_{1}}\int_{\mathcal{M}_{t}}
X_{\b}\na_{\a}S^{\a\b}d\mu+\int_{t_{0}}^{t_{1}}\int_{\mathcal{M}_{t}}
T^{\a\b}(\c)\na_{(\a}X_{\b)}d\mu
\ee
and therefore using Stokes' theorem and the scalar-tensor equation (\ref{st}) we
obtain
\bq
E_{t_{1}}(\c )-E_{t_{0}}(\c )
&=&
\int_{\mathcal{M}_{t_{1}}}S^{\a\b}X_{\b}N_{\a}d\mu_{t_{1}}
-\int_{\mathcal{M}_{t_{0}}}S^{\a\b}X_{\b}N_{\a}d\mu_{t_{0}}\nonumber\\
&-&
\int_{t_{0}}^{t_{1}}\int_{\mathcal{M}_{t}}
S^{\a\b}\na_{\a}X_{\b} +\frac{1}{2}\int_{t_{0}}^{t_{1}}\int_{\mathcal{M}_{t}}
(S^{\a\b}-T^{\a\b}(\psi))(\na_{\a}X_{\b}+\na_{\b}X_{\a})\nonumber\\
&=&
\int_{\mathcal{M}_{t_{1}}}S^{\a\b}X_{\b}N_{\a}d\mu_{t_{1}}
-\int_{\mathcal{M}_{t_{0}}}S^{\a\b}X_{\b}N_{\a}d\mu_{t_{0}}\nonumber\\
&-&
\frac{1}{2}\int_{t_{0}}^{t_{1}}\int_{\mathcal{M}_{t}}
S^{\a\b}(\na_{\a}X_{\b}-\na_{\b}X_{\a})\nonumber\\
&-&
\frac{1}{2}\int_{t_{0}}^{t_{1}}\int_{\mathcal{M}_{t}}
T^{\a\b}(\psi)(\na_{\a}X_{\b}+\na_{\b}X_{\a})\nonumber\\
&=&
\int_{\mathcal{M}_{t_{1}}}S^{\a\b}X_{\b}N_{\a}d\mu_{t_{1}}
-\int_{\mathcal{M}_{t_{0}}}S^{\a\b}X_{\b}N_{\a}d\mu_{t_{0}}\nonumber\\
&-&
\int_{t_{0}}^{t_{1}}\int_{\mathcal{M}_{t}}
S^{\a\b}\na_{[\a}X_{\b ]}-\int_{t_{0}}^{t_{1}}\int_{\mathcal{M}_{t}}
T^{\a\b}(\psi)\na_{(\a}X_{\b )}.
\eq
The penultimate term in the last equality of this equation is zero, as the first tensor
in the product is symmetric and the second antisymmetric, and so we
find that
\be
E_{t_{1}}(\c )-E_{t_{0}}(\c )=\int_{\mathcal{M}_{t_{1}}}S^{\a\b}X_{\b}N_{\a}d\mu_{t_{1}}
-\int_{\mathcal{M}_{t_{0}}}S^{\a\b}X_{\b}N_{\a}d\mu_{t_{0}}-
\int_{t_{0}}^{t_{1}}\int_{\mathcal{M}_{t}}
T^{\a\b}(\psi)\na_{(\a}X_{\b )}.
\ee
Now, since $\na_\a S^{\a\b}=G^{\a\b}\na_{\a}\f$, we find that the
first two terms can be expressed more simply as follows
\be
\int_{\mathcal{M}_{t_{1}}}S^{\a\b}X_{\b}N_{\a}d\mu_{t_{1}}
-\int_{\mathcal{M}_{t_{0}}}S^{\a\b}X_{\b}N_{\a}d\mu_{t_{0}}=
\int_{t_{0}}^{t_{1}}\int_{\mathcal{M}_{t}}S^{\a\b}\na_{\a}X_{\b}+
\int_{t_{0}}^{t_{1}}\int_{\mathcal{M}_{t}}G^{\a\b}X_{\b}\pa_{\a}\chi .
\ee
Therefore using Eq. (\ref{12}) we find that
\beq\label{st-e3}
E_{t_{1}}(\chi+\psi)-E_{t_{0}}(\chi+\psi)&=&\int_{t_{0}}^{t_{1}}\int_{\mathcal{M}_{t}}
\chi G^{\a\b}\na_\a X_\b+\int_{t_{0}}^{t_{1}}\int_{\mathcal{M}_{t}}G^{\a\b}X_\b\pa_\a\chi \nonumber\\
&=&\int_{t_{0}}^{t_{1}}\int_{\mathcal{M}_{t}}G^{\a\b}(\chi\na_\a X_\b +X_\b\pa_\a\chi
)\nonumber\\
&=&\int_{t_{0}}^{t_{1}}\int_{\mathcal{M}_{t}}G^{\a\b}\na_\a (\chi X_\b
)
\eeq
and we are led to the following result.
\begin{theorem}
The total slice energy of the dilaton-matter system  satisfying the scalar-tensor
equations (\ref{st}) is given by
\be\label{st-e2}
E_{t_{1}}(\chi+\psi)-E_{t_{0}}(\chi+\psi)=
\int_{\mathcal{M}_{t_{1}}}S^{\a\b} X_\a N_\b d\mu_{t_{1}}-
\int_{\mathcal{M}_{t_{0}}}S^{\a\b} X_\a N_\b d\mu_{t_{0}}.
\ee
\end{theorem}

\section{Discussion}
In conclusion we have found the different forms that slice energy
takes in various classes of generalized theories of gravitation
which include higher-order gravity theories and scalar-tensor ones.
These forms may be described symbolically as follows:
\be\label{last}
E_{t_{1}}(\lambda+\psi)-E_{t_{0}}(\lambda+\psi)=
\int_{\mathcal{M}_{t_{1}}}\Lambda^{\a\b}X_\a N_\b d\mu_{t_{1}}-
\int_{\mathcal{M}_{t_{0}}}\Lambda^{\a\b} X_\a N_\b d\mu_{t_{0}}.
\ee
Here $\lambda$ denotes either an inflaton field, $\f$, which
couples to matter in general relativity or in  the Einstein frame
in higher-order gravity,  the scalar field $\xi$ which may
appear in the Jordan frame of higher-order gravity, or the dilaton
$\chi$ in scalar-tensor theory, while $\Lambda$ is a gravitational
operator defining the left-hand-sides of the associated field equations,
that is, $\Lambda$ is $G^{\a\b}, \tilde{G}^{\a\b}, L^{\a\b}$ or $S^{\a\b}$
respectively.

The above analysis allows for some conclusions to be
drawn concerning the scalar fields present in the metric theories considered so far.
Firstly Eq. (\ref{last}) gives a universal formula for the scalar
field energies appearing in different metric theories of gravitation and
may provide an answer to the issue discussed in the Introduction and
first posed by Brans in \cite{brans}
of the relative behaviour of these fields. According to the results of this paper we may
conclude that an observer who
moves from slice to slice in spacetime finds that the total slice
energy of the system composed of a scalar field and a
matterfield depends solely on the tensorfield $\Lambda$ which defines the
gravitational sector of the theory. We may therefore use Eq.
(\ref{last}) as \emph{a definition} of the scalar field we wish to
couple in a given theory and conclude that this equation provides a clue to the relative
differences of the scalar fields discussed above.

Secondly it follows that in all these conformally related frames
slice energy maintains the same structural definition, namely, Eq.
(\ref{last}), and its conservation properties are structurally the
same in each conformal frame. Although its form obviously depends
on the particular theory considered, as in Eqs. (\ref{gr-e1}),
(\ref{hog-e1}), (\ref{hog-e3}) and (\ref{st-e3}), its meaning and
basic properties do not depend upon the particular conformal frame
representation used. In this sense it may be called an invariant.
This is true despite the existence of several different `measures
of physical inequivalence' sometimes used in the literature to
show an inequivalence of the conformally related frames,  such as
conservation laws, positivity of energy, the existence of a stable
ground state etc. Here we have a quantity of obvious physical
meaning the properties of which are in the above sense insensitive
to conformal transformations. We may therefore conclude that with
respect to slice energy these different conformal frames are
physically equivalent and this conclusion reinforces
that reached in \cite{last}.

\section*{Acknowledgements}
We thank Peter leach for useful comments on the manuscript.

\end{document}